\documentclass[a4paper]{article}
\usepackage[utf8]{inputenc}
\usepackage[T1]{fontenc}

\usepackage[left=15mm,right=15mm,bottom=15mm,top=15mm]{geometry}

\usepackage{longtable}
\usepackage{booktabs}
\usepackage{microtype}
\usepackage[hyphens]{url}
\usepackage{dblfnote}
\DFNalwaysdouble%

\usepackage{stmaryrd}                         % Short arrow
\usepackage{textcomp}                         % Upquote=true
\usepackage{xcolor}                           % color for the postbreak character
\usepackage{listings}
\usepackage{natbib}
\usepackage[hidelinks]{hyperref}
\usepackage[nameinlink,capitalize]{cleveref}
\makeatletter
\newcommand\cref@smugglelabel{\let\cref@currentlabel\cref@gcurrentlabel@temp}
\newcommand\cref@updatelabeldata[1]{%
	\cref@constructprefix{#1}{\cref@result}%
	\@ifundefined{cref@#1@alias}%
	{\def\@tempa{#1}}%
	{\def\@tempa{\csname cref@#1@alias\endcsname}}%
	\protected@xdef\cref@gcurrentlabel@temp{%
		[\@tempa][\arabic{#1}][\cref@result]%
	\csname p@#1\endcsname\csname the#1\endcsname}%
	\aftergroup\cref@smugglelabel%
}
\AddToHook{label}{\cref@updatelabeldata{\@currentcounter}} 
\makeatother

\usepackage{lstlangclean}
\usepackage{lstlanghaskell}
\newcommand{\cleaninline}[1]{\lstinline[language=Clean,postbreak=]|#1|}
\newcommand{\haskellinline}[1]{\lstinline[language=Haskell,style=haskell,postbreak=]|#1|}
\lstnewenvironment{lstClean}[1][]
	{% 
		\lstset{language=Clean, #1}
		
	}
	{}
\lstnewenvironment{lstCleanT}[1][]
	{% 
		\lstset{language=Clean,aboveskip=0pt,belowskip=-1\baselineskip,frame=,#1}
		\vspace*{-.5\baselineskip}
	}
	{}
\lstnewenvironment{lstHaskell}[1][]
	{%
		\lstset{language=Haskell,style=haskell,#1}%
		
	}
	{}
\lstnewenvironment{lstHaskellT}[1][]
	{%
		\lstset{language=Haskell,style=haskell,aboveskip=0pt,belowskip=-2\baselineskip,#1}%
		\vspace*{-.8\baselineskip}
	}
	{}

\newcommand{\Haskell}[0]{Haskell}
\newcommand{\Clean}[0]{Clean}
\newcommand{\GenericHaskell}[0]{Generic H$\forall$skell}

\newcommand{\GHCmod}[1]{\textsf{#1}}
\newcommand{\requiresGHCmod}[1]{\footnote{Requires \GHCmod{#1} to be enabled.}}

\newcommand{\subject}[1]{%
	\midrule
	\multicolumn{2}{c}{#1}\\
	\midrule
}

\author{%
		Mart Lubbers\\
		\texttt{mart@cs.ru.nl}
	\and
		Peter Achten\\
		\texttt{p.achten@cs.ru.nl}
}
\title{Clean for Haskell Programmers}
\date{\today}

\begin{document}
\maketitle

This note is meant to give people who are familiar with the functional programming language \Haskell{}\footnote{
By \Haskell{} we mean GHC's \Haskell{}}
a concise overview of \Clean{}\footnote{
By \Clean{} we mean \Clean{} 3.1 (\url{https://clean-lang.org}).}
language elements and how they differ from \Haskell{}.
Many of this is based on work by \citet{achten_clean_2007} although that was based on \Clean{} 2.1 and Haskell98.
Obviously, this summary is not exhaustive, a complete specification of the \Clean{} language can be found in the latest language report \citep{plasmeijer_clean_2021}.
The main goal is to support the reader when reading \Clean{} code.
\Cref{tbl:syn_clean_haskell} shows frequently occurring \Clean{} language elements on the left side and their \Haskell{} equivalent on the right side.
Other \Clean{} language constructs that also frequently occur in \Clean{} programs, but that do not appear in the table are:

\paragraph{Modules}
\Clean{} modules have separate definition (headers) and implementation files.
The definition module contains the class definitions, instances, function types and type definitions (possibly abstract).
Implementation modules contain the function implementations as well.
This means that only what is defined in the definition module is exported in \Clean{}.
This differs greatly from \Haskell{}, as there is only a module file there.
Choosing what is exported in \Haskell{} is done using the \haskellinline{module Mod(...)} syntax.

\paragraph{Strictness}
In \Clean{}, by default, all expressions are evaluated lazily.
Types can be annotated with a strictness attribute (\cleaninline{!}), resulting in the values being evaluated to head-normal form before the function is entered.
In \Haskell{}, in patterns, strictness can be enforced using \haskellinline{!}\requiresGHCmod{BangPatterns}.
Within functions the strict let (\cleaninline{#!}) can be used to force evaluate an expression, in \Haskell{} \haskellinline{seq} or \haskellinline{\$!} is used for this.

\paragraph{Uniqueness typing}
Types in \Clean{} may be \emph{unique}, which means that they may not be shared \citep{barendsen_uniqueness_1996}.
The uniqueness type system allows the compiler to generate efficient code because unique data structures can be destructively updated.
Furthermore, uniqueness typing serves as a model for side effects as well.
Clean uses the \emph{world-as-value} paradigm where \cleaninline{World} represents the external environment and is always unique.
A program with side effects is characterised by a \cleaninline{Start :: *World -> *World} start function.
In \Haskell{}, interaction with the world is done using the \haskellinline{IO} monad.
The \haskellinline{IO} monad could very well be---and actually is---implemented in \Clean{} using a state monad with the \cleaninline{World} as a state.
Besides marking types as unique, it is also possible to mark them with uniqueness attributes variables \cleaninline{u:} and define constraints on them.
For example, to make sure that an argument of a function is at least as unique as another argument.
Finally, using \cleaninline{.} (a dot), it is possible to state that several variables are equally unique.
Uniqueness is propagated automatically in function types but must be marked manually in data types.
Examples can be seen in \cref{lst:unique_examples}.

\begin{lstClean}[label={lst:unique_examples},caption={Examples of uniqueness annotations in \Clean{}.}]
f :: (Int, *World) -> *World // uniqueness is propagated automatically for function types (i.e. *(Int, *World)))
f :: *a -> *a                // f works on unique values only
f :: .a -> .a                // f works on unique and non-unique values
f :: v:a u:b -> u:b, [v<=u]  // f works when a is less unique than b
\end{lstClean}

\paragraph{Generics}
Generic functions \citep{jeuring_polytypic_1996}---otherwise known as polytypic or kind-indexed fuctions---are built into \Clean{} \citet[Chp.~7.1]{plasmeijer_clean_2021}\citep{alimarine_generic_2005} whereas in \Haskell{} they are implemented as a library~\citep[Chp.~6.19.1]{ghc_team_ghc_2021}.
The implementation of generics in \Clean{} is very similar to that of \GenericHaskell{}~\citep{hinze_generic_2003}.
Metadata about the types is available using the \cleaninline{of} syntax, giving the function access to metadata records.
This abundance of metadata allows for very complex generic functions that near the expression level of template metaprogramming.
\Cref{lst:generic_eq} shows an example of a generic equality and \cref{lst:generic_print} of a generic print function utilising the metadata.

\paragraph{GADTs}
Generalised algebraic data types (GADTs) are enriched data types that allow the type instantiation of the constructor to be explicitly defined \citep{cheney_first-class_2003,hinze_fun_2003}.
While GADTs are not natively supported in \Clean{}, they can be simulated using embedding-projection pairs or equivalence types~\citep[Sec.~2.2]{cheney_lightweight_2002}.
To illustrate this, \cref{lst:gadt_clean,lst:gadt_haskell} show an example GADT implemented in \Clean{} and \Haskell{}\requiresGHCmod{GADTs} respectively.

%\newpage
%\section{Syntax}
\small
\begin{longtable}[c]{p{.425\linewidth}p{.525\linewidth}}
	\caption[]{Syntactical differences between \Clean{} and \Haskell{}.}%
	\label{tbl:syn_clean_haskell}\\
	\toprule
	\Clean{} & \Haskell{}\\
	\endfirsthead%
%	\caption[]{(continued)}\\
	\toprule
%	\Clean{} & \Haskell{}\\
%	\midrule
	\endhead%

	\subject{Comments}
	\cleaninline{// single line} & \haskellinline{-- single line}\\
	\cleaninline{/* multi line /* nested */ */} & \haskellinline{\{- multi line \{- nested -\} \}}\\

	\subject{Imports}
	\begin{lstCleanT}
import Mod0
import Mod1 => qualified f, :: t
	\end{lstCleanT}
	&
	\begin{lstHaskellT}
import Mod0 (f, t)
import qualified Mod1 (f, t)
import Mod1 hiding (f, t)

	\end{lstHaskellT}
	\\

	\subject{Basic types}
	\cleaninline{42 :: Int} & \haskellinline{42 :: Int}\\
	\cleaninline{True :: Bool} & \haskellinline{True :: Bool}\\
	\cleaninline{toInteger 42 :: Integer} & \haskellinline{42 :: Integer}\\
	\cleaninline{38.0 :: Real} & \haskellinline{38.0 :: Float -- or Double}\\
	\cleaninline{\"Hello\" +++ \"World\" :: String}\footnote{Strings are unboxed character arrays.}
	& \haskellinline{\"Hello\" ++ \"World\" :: String}\footnote{Strings are lists of characters or overloaded if \GHCmod{OverloadedStrings} is enabled.}\\
	\cleaninline{['Hello'] :: [Char]} & \haskellinline{\"Hello\" :: String}\\
	\cleaninline{?t} & \haskellinline{Maybe t}\\
	\cleaninline{(?None, ?Just e)} & \haskellinline{(Nothing, Just e)}\\

	\subject{Type definitions}
	\cleaninline{:: T a0 ... :== t} & \haskellinline{type T a0 ... = t}\\
	\cleaninline{:: T a0 ... = C0 f0 f1 ... \| C1 f0 f1 ... \| ...} & \haskellinline{data T a0 ... = C0 f0 f1 ... \| C1 f0 f1 ... \| ...}\\
	\cleaninline{:: T a0 ... = \{ f0 :: t0, f1 :: t1, ... \}} & \haskellinline{data T a0 ... = T \{ f0 :: t0, f1 :: t1, ... \} }\\
	\cleaninline{:: T a0 ... =: t} & \haskellinline{newtype T a0 ... = t}\\
	\cleaninline{:: T = E.t: Box t \& C t} & \haskellinline{data T = forall t.C t => Box t}\requiresGHCmod{ExistentialQuantification}\\

	\subject{Function types}
	\cleaninline{f0 :: a0 a1 ... -> t \| C0 v0 \& C1, C2 v1}
		& \haskellinline{f0 :: (C0 v0, C1 v1, C2 v2) => a0 -> a1 ... -> t}\\
	\begin{lstCleanT}
(+) infixl 6 :: Int Int -> Int
	\end{lstCleanT}
	&
		\begin{lstHaskellT}
infixl 6 +
(+) :: Int -> Int -> Int

		\end{lstHaskellT}
		\\
	\begin{lstCleanT}
qid :: (A.a: a -> a) -> (Bool, Int)
qid id = (id True, id 42)
	\end{lstCleanT}
		&
	\begin{lstHaskellT}
qid(*\requiresGHCmod{RankNTypes}*) :: (forall a: a -> a) -> (Bool, Int)
qid id = (id True, id 42)
	\end{lstHaskellT}
	\\

	\subject{Type classes}
	\cleaninline{class f a :: t} & \haskellinline{class F a where f :: t}\\
	\cleaninline{class C a \| C0, ... , Cn a} & \haskellinline{class (C0 a, ..., Cn, a) => C a}\\
	\cleaninline{class C s ~m where ...} & \haskellinline{class C s m \| m -> s where ...}\requiresGHCmod{MultiParamTypeClasses}\\
	\cleaninline{instance C t \| C0 t \& C1 t ...  where ...} & \haskellinline{instance (C0 a, C1 a, ...) => C t where ...}\\

	\subject{As pattern}
	\cleaninline{x=:p}\footnote{May also be used as a predicate, e.g.\ \cleaninline{if (e0 =: []) e1 e2}.} & \haskellinline{x@p}\\

	\subject{Lists}
	\cleaninline{[1,2,3]} & \haskellinline{[1,2,3]}\\
	\cleaninline{[x:xs]} & \haskellinline{x:xs}\\
	\cleaninline{[e \\\\ e<-xs \| p e]} & \haskellinline{[e \| e<-xs, p e]}\\
	\cleaninline{[l \\\\ l<-xs, r<-ys]} & \haskellinline{[l \| l<-xs, r<-ys]}\\
	\cleaninline{[(l, r) \\\\ l<-xs \& r<-ys]} & \haskellinline{[(l, r) \| (l, r)<-zip xs ys]} or \haskellinline{[(l, r) \| l<-xs \| r<-ys]}\requiresGHCmod{ParallelListComp}\\

	\subject{Lambda expressions}
	\cleaninline{\\a0 a1 ...->e} or \cleaninline{\\....e} or \cleaninline{\\...=e} & \haskellinline{\\a0 a1 ...->e}\\

	\subject{Case distinction}
	\cleaninline{if p e0 e1} & \haskellinline{if p then e0 else e1}\\
	\begin{lstCleanT}
case e of
	p0 -> e0 // or \texttt{p\textsubscript{0} = e\textsubscript{0}}
	...
	\end{lstCleanT}
	&
	\begin{lstCleanT}
case e of
	p0 -> e0
	...
	\end{lstCleanT}
	\\
	\begin{lstCleanT}
f p0 p1 ...
	| c         = t
	| otherwise = t // or \texttt{= t}
	\end{lstCleanT}
	&
	\begin{lstHaskellT}
f p0 p1 ...
	| c         = t
	| otherwise = t
	\end{lstHaskellT}
	\\

	\subject{Records}
	\cleaninline{:: R = \{ f :: t \}} & \haskellinline{data R = R \{ f :: t \}}\\
	\cleaninline{r = \{ f = e \}} & \haskellinline{r = R \{e\}}\\
	\cleaninline{r.f} & \haskellinline{f r}\\
	\cleaninline{r!f}\footnote{Field selection from unique records.} & \haskellinline{(\\v->(f v, v)) r}\\
	\cleaninline{\{r \& f = e \}} & \haskellinline{r \{ f = e \}}\\

	\subject{Record patterns}
	\cleaninline{:: R0 = \{ f0 :: R1 \}} & \haskellinline{data R0 = R0 \{ f0 :: R1 \}}\\
	\cleaninline{:: R1 = \{ f1 :: t \}} & \haskellinline{data R1 = R1 \{ f1 :: t \}}\\
	\cleaninline{g \{ f0 \} = e f0} & \haskellinline{g (R0 \{f0=x\}) = e x} or \haskellinline{g (R0 \{f0\}) = e f0}\requiresGHCmod{RecordPuns}\\
	\cleaninline{g \{ f0 = \{f1\} \} = e f1} & \haskellinline{g (R0 \{f0=R1 \{f1=x\}\}) = e x}\\

	\subject{Arrays}
	\cleaninline{:: A :== \{t\}} & \haskellinline{type A = Array Int t}\\
	\cleaninline{a = \{v0, v1, ...\}} & \haskellinline{array (0, n+1) [(0, v0), (1, v1), ..., (n, ...)]}\\
	\cleaninline{a = \{e \\\\ p <-: a\}} & \haskellinline{array (0, length a-1) [e \| (i, p) <- zip [0..] a]}\\
	\cleaninline{a.[i]} & \haskellinline{a!i}\\
	\cleaninline{a![i]}\footnote{Field selection from unique arrays.} & \haskellinline{(\\v->(v!i, v)) a}\\
	\cleaninline{\{ a \& [i] = e\}} & \haskellinline{a//[(i, e)]}\\

	\subject{Dynamics}
	\begin{lstCleanT}
f :: a -> Dynamic | TC a
f e = dynamic e
	\end{lstCleanT}
	&
	\begin{lstHaskellT}
f :: Typeable a => a -> Dynamic
f e = toDyn e
	\end{lstHaskellT}
	\\
	\begin{lstCleanT}
g :: Dynamic -> t
g (e :: t) = e0
g  e      = e1
	\end{lstCleanT}
	&
	\begin{lstHaskellT}
g :: Dynamic -> t
g d = case fromDynamic d of
	Just e  -> e0
	Nothing -> e1

	\end{lstHaskellT}
	\\
	
	\subject{Function definitions}
	\begin{lstCleanT}
f p0
	# q0 = e0
	= e
	\end{lstCleanT}
	&
	\begin{lstHaskellT}
f p0
    = e(*$[x := x']$*)
  where q0(*$[x := x']$*) = e0 -- \textit{for each }$x \in \mathit{var}(q_0) \cap \mathit{var}(e_0)$
	\end{lstHaskellT}
	\\
	\bottomrule
\end{longtable}
\normalsize

\newpage
\noindent%
\begin{minipage}[b]{.5\textwidth}
\begin{lstlisting}[language=Clean,firstline=4,lastline=24,label={lst:gadt_clean},caption={Expression GADT in \Clean{}.}]
:: BM a b = { ab :: a -> b, ba  :: b -> a }
bm :: BM a a
bm = {ab=id, ba=id}

:: Expr a
	= E.e: Lit (BM a e) e & toString e
	| E.e: Add (BM a e)    (Expr e) (Expr e) & + e
	| E.e: Eq  (BM a Bool) (Expr e) (Expr e) & == e
lit e = Lit bm e
add l r = Add bm l r
eq l r = Eq bm l r

eval :: (Expr a) -> a
eval (Lit bm e)   = bm.ba e
eval (Add bm l r) = bm.ba (eval l + eval r)
eval (Eq  bm l r) = bm.ba (eval l == eval r)

print :: (Expr a) -> String
print (Lit _ e)   = toString e
print (Add _ l r) = print l +++ "+" +++ print r
print (Eq  _ l r) = print l +++ "==" +++ print r
\end{lstlisting}
\end{minipage}
\begin{minipage}[b]{.45\textwidth}
\begin{lstlisting}[language=Haskell,style=haskell,firstline=4,label={lst:gadt_haskell},caption={Expression GADT in \Haskell{}.}]
data Expr a where
    Lit :: Show a => a -> Expr a
    Add :: Num a  => Expr a -> Expr a -> Expr a
    Eq  :: Eq e   => Expr e -> Expr e -> Expr Bool

eval :: Expr a -> a
eval (Lit e)   = e
eval (Add l r) = eval l + eval r
eval (Eq  l r) = eval l == eval r

print :: Expr a -> String
print (Lit e)   = show e
print (Add l r) = print l ++ "+"  ++ print r
print (Eq  l r) = print l ++ "==" ++ print r
\end{lstlisting}
\end{minipage}

\begin{lstlisting}[language=Clean,firstline=4,label={lst:generic_eq},caption={Generic equality function in \Clean{}.}.]
generic gEq a :: a a -> Bool

gEq{|Int|}           x            y           = x == y
gEq{|Bool|}          x            y           = x == y
gEq{|Real|}          x            y           = x == y
gEq{|Char|}          x            y           = x == y
gEq{|UNIT|}          x            y           = True
gEq{|OBJECT|} f     (OBJECT x)   (OBJECT y)   = f x y
gEq{|CONS|}   f     (CONS x)     (CONS y)     = f x y
gEq{|RECORD|} f     (RECORD x)   (RECORD y)   = f x y
gEq{|FIELD|}  f     (FIELD x)    (FIELD y)    = f x y
gEq{|PAIR|}   fl fr (PAIR lx rx) (PAIR ly ry) = fl lx ly && fr rx ry
gEq{|EITHER|} fl _  (LEFT x)     (LEFT y)     = fl x y
gEq{|EITHER|} _  fr (RIGHT x)    (RIGHT y)    = fr x y
gEq{|EITHER|} _  _  _            _            = False

:: T = C1 Int ([Char], ?Bool) | C2
derive gEq [], T, (,), ?

Start = (gEq{|*|} C2 (C1 42 ([], ?Just True)), gEq{|*->*|} (<) [1,2,3] [2,3,4])
// (False, True)
\end{lstlisting}
\begin{lstlisting}[language=Clean,firstline=4,label={lst:generic_print},caption={Generic print function in \Clean{}.}]
generic gPrint a :: a [String] -> [String]

gPrint{|Int|}           x         acc = [toString x:acc]
gPrint{|Bool|}          x         acc = [toString x:acc]
gPrint{|Real|}          x         acc = [toString x:acc]
gPrint{|Char|}          x         acc = [toString x:acc]
gPrint{|UNIT|}          x         acc = acc
gPrint{|PAIR|}   fl fr (PAIR l r) acc = fl l [" ":fr r acc]
gPrint{|EITHER|} fl _  (LEFT x)   acc = fl x acc
gPrint{|EITHER|} _  fr (RIGHT x)  acc = fr x acc

gPrint{|OBJECT|}        f     (OBJECT x) acc = f x acc
gPrint{|CONS of gcd|}   f     (CONS x)   acc = ["(", gcd.gcd_name, " ":f x [")":acc]]
gPrint{|RECORD of grd|} f     (RECORD x) acc = ["{", grd.grd_name, " | ":f x ["}":acc]]
gPrint{|FIELD of gfd|}  f     (FIELD x)  acc = [pre, gfd.gfd_name, "=":f x acc]
where
	pre = if (gfd.gfd_index == 0) "" ", "

:: T = {f1 :: Int, f2 :: (Real, [?Int])}
derive gPrint (,), [], ?, T

Start = gPrint{|*|} {f1=42, f2=(3.14, [?None])} []
// \verb#{T | f1=42 , f2=(_Tuple2 3.14 (_Cons (_!None ) (_Nil )))}#
\end{lstlisting}

\bibliographystyle{plainnat}
\bibliography{refs}
\end{document}